\documentclass[aps,prb,showpacs,twocolumn,superscriptaddress,10pt,footinbib]{revtex4}

\usepackage{graphicx}
\usepackage{epstopdf}
\usepackage{amsmath}
\usepackage{amssymb}
\usepackage{amsfonts}
\usepackage{latexsym}
\usepackage{color}

\usepackage[colorlinks,bookmarks=false,citecolor=blue,linkcolor=red,urlcolor=blue]{hyperref}
\usepackage{verbatim}

\def\be{\begin{equation}}
\def\ee{\end{equation}}
\def\bea{\begin{eqnarray}}
\def\eea{\end{eqnarray}}

\def\vec{\mathbf}
\def\mc{\mathcal}

\begin{document}
\title{Theory of severe slowdown in the relaxation of rings and clusters with antiferromagnetic interactions}
\author{Ioannis Rousochatzakis}\email{ioannis.rousochatzakis@epfl.ch}
\affiliation{Institut de th\'eorie des ph\'enom\`enes physiques, Ecole polytechnique f\'ed\'erale de Lausanne, CH-1015 Lausanne, Switzerland}
\author{Andreas L\"auchli}
\affiliation{Max Planck Institut f\"ur Physik komplexer Systeme, D-01187 Dresden, Germany}
\author{Ferdinando Borsa}
\affiliation{Dipartimento di Fisica ``A. Volta'', e Unit\`a CNISM, Universit\`a di Pavia, I-27100 Pavia, Italy}
\affiliation{Ames Laboratory and Department of Physics and Astronomy, Iowa State University, Ames, Iowa, 50011}
\author{Marshall Luban}
\affiliation{Ames Laboratory and Department of Physics and Astronomy, Iowa State University, Ames, Iowa, 50011}

\begin{abstract}
We show that in the severe slowing down temperature regime the relaxation 
of antiferromagnetic rings and similar magnetic nanoclusters is governed by the quasi-continuum portion 
of their quadrupolar fluctuation spectrum and not by the lowest excitation lines. 
This is at the heart of the intriguing near-universal power-law temperature dependence of the electronic correlation frequency $\omega_c$
with an exponent close to 4. The onset of this behavior is defined by an energy scale which is fixed by the lowest spin gap $\Delta_0$. 
This explains why experimental curves of $\omega_c$ for different cluster sizes and spins nearly coincide when $T$ is rescaled by $\Delta_0$. 
\end{abstract}


\maketitle

\section{Introduction}
A central issue for the control and manipulation of electronic spin degrees of freedom 
in the field of molecular nanomagnetism\cite{GSV} and related areas  
is the understanding and characterization of the various microscopic mechanisms of relaxation and decoherence
which stem from the interactions with the underlying degrees of freedom of the host lattice. 
It is by now well established that in the large majority of magnetic clusters, the relaxation of the magnetization shows a dramatic slowing down 
which sets in already at relatively high temperatures $T$ 
and is characterized by a single electronic frequency cutoff $\omega_c$\cite{GSV,NMR_Santini,Rousochatzakis_PRB}. 
While the majority of freezing mechanisms reported in the literature 
[e.g., onsite anisotropy in single molecule magnets\cite{GSV}, anisotropy and critical fluctuations in single chain magnets\cite{GSV}, 
phonon trapping in Ni$_{10}$\cite{Ni10}], have been understood to a large extent on the basis of the nature of the lowest excitations,
the case of antiferromagnetic rings (AFMR's) has been exceptional and very intriguing. 
Indeed, as shown by Nuclear Magnetic Resonance (NMR) experiments by Baek \textit{et al.}\cite{NMR_Baek}, in the 
regime where the dramatic slowing down effect takes place, 
$\omega_c$ follows a near-universal power-law $T$ dependence with an exponent close to 4, a behavior clearly inconsistent with 
a relaxation scenario based on the lowest excitation lines. 
It is also striking that the experimental curves of $\omega_c$ from various magnetic rings with different sizes and spins $s>1/2$
nearly coincide when $T$ is rescaled by the lowest spin gap $\Delta_0$. 

On the theoretical side, the first microscopic calculation of $\omega_c$ in AFMR's has been given by Santini \textit{et al.}\cite{NMR_Santini} 
who conclude on this problem that the relevant $T$ regime is too narrow to substantiate any power-law scaling 
and that instead $\omega_c$ is approximately exponential in $1/T$. 

\begin{figure}[!b]
\centering
\includegraphics*[width=0.9\linewidth]{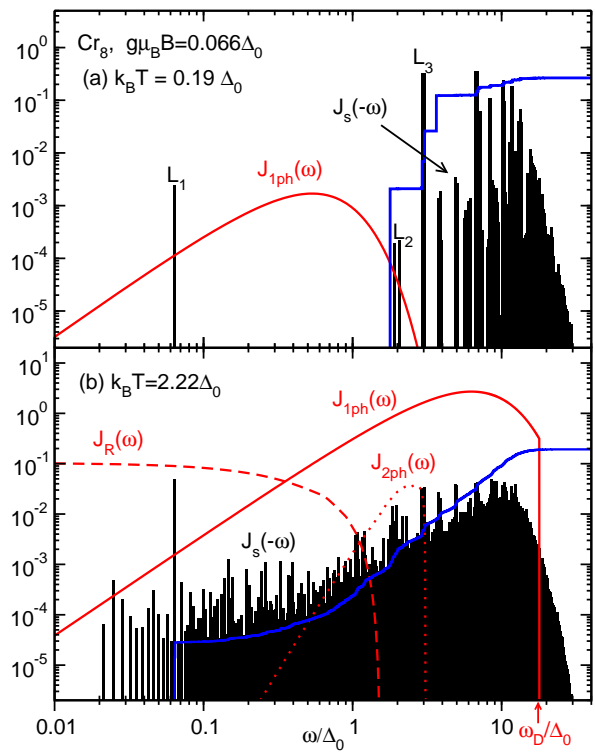}
\caption{(Color online) The origin of the dramatic slowing down effect in AFMR's, demonstrated here for the Cr$_8$ cluster 
at $k_B T=0.19 \Delta_0$ (a) and $2.22 \Delta_0$ (b), with $\hbar\omega_e=0.066 \Delta_0$ and $\hbar\omega_D=10 J$.  
The correlation frequency $\omega_c$ is proportional [cf. Eq.~(\ref{wc1.Eq}) or (\ref{wc2.Eq})] to the overlap between 
the quadrupolar spin density $J_s(-\omega)$ (series of $\delta$-peaks, here shown with a logarithmic mesh in $\omega$) and the phononic density  
$J_{\overline{\Phi}^\dagger \overline{\Phi} }(\omega)$ which includes one-phonon (solid red), two-phonon (dotted red) and Raman (dashed red) processes. We also show (solid blue line) the integrated density $\int_0^{\omega}d\omega'J_s(-\omega')$.
The lowest three groups of excitation lines at low T are denoted by L$_1$, L$_2$ and L$_3$ (cf. text).}  
\label{Overlap.Fig}
\end{figure}

Here we present a new microscopic theory which supports the scaling hypothesis of Baek {\it et al.}\cite{NMR_Baek} 
and resolves its physical origin in a clear and transparent way. 
Our central key finding is that, in the severe slowing down $T$ regime, the relaxation is 
not governed by the lowest excitation lines of the quadrupolar spectral density but by a forest of quasi-continuum excitations
at higher energies (cf. Fig.~\ref{Overlap.Fig}).  
This is at the heart of the power-law temperature dependence of $\omega_c$ with an exponent which is close to 4 (cf. Eq.~(\ref{wc4.Eq})). 
The onset of this behavior is defined by an energy scale which is fixed by the lowest spin gap $\Delta_0$, 
and this explains why curves from different rings fall almost on top of each other when $T$ is rescaled by $\Delta_0$. 
The present theory is corroborated by a model calculation of the nuclear spin-lattice relaxation rate $1/T_1$ 
which includes one-phonon, two-phonon as well as Raman processes 
and shows excellent agreement with experimental data for the Cr$_8$, Fe$_6$Li and Fe$_6$Na clusters.

Our theory is based on the formalism developed in Ref. [\onlinecite{Rousochatzakis_PRB}].
The advantage of this method is that it is based on an analytical formula which gives $\omega_c$ in terms of a frequency overlap 
between an electronic and a phononic spectral density function [cf. Eq.~(\ref{wc1.Eq}) or (\ref{wc2.Eq})  below]. 
Thus by monitoring the variation of these spectra with $T$ and $\omega$ one is able to identify the dominant relaxation channels 
in each $T$ regime of interest. 
The central approximation for the derivation of this analytical formula of $\omega_c$ is that the spin Hamiltonian $\mc{H}_0$ 
commutes with the total magnetization $S_z$ of the cluster. For the majority of magnetic clusters reported in the literature this 
approximation is an excellent one for the study of relaxation phenomena at not too low T ($T\gtrsim 1$  K). 
Indeed, the dominant energy scale in the problem is set by the isotropic Heisenberg exchange and very often by a uniaxial onsite anisotropy. 
There exist several types of anisotropic interactions which do not conserve $S_z$, but these are typically very small ($\lesssim 1$ K) 
and thus play a significant role only at very low T. Thus, although the present work is devoted to the slowing down mechanism in AFMR's 
the method presented here can be applied to the majority of magnetic clusters reported in the literature. 
In particular, we will argue that the slowing down mechanism in AFMR's must be common
in the general class of antiferromagnetic nanomagnets (which includes e.g., grids and other clusters) 
due to their very similar spectral structure (cf. below).

\section{Method and Model}
We consider a magnetic ring with an even number $N$ of spins $s>1/2$\cite{note0} described by the Hamiltonian
\be
\mc{H}_0= J\sum_i \vec{s}_i \cdot \vec{s}_{i+1} + g \mu_B B S_z~, 
\ee
with periodic boundary conditions. The first term describes the antiferromagnetic ($J>0$) exchange between nearest-neighbor spins, 
the second is the Zeeman energy in a field $\vec{B}=B\vec{e}_z$, $\vec{S}=\sum_i \vec{s}_i$ is the total spin, 
$g\simeq 2$, and $\mu_B$ is the Bohr magneton. 
It is known\cite{SchnackLuban,Waldmann,Larry} that the energy spectrum of $\mc{H}_0$ is bounded from below by an 
excitation band $E_S \simeq \Delta_0 S (S+1)/2$, where $\Delta_0\simeq 4 J/N$ (cf. Ref. [\onlinecite{Larry}] for the validity range of this scaling) 
is the lowest singlet-triplet gap. 
At higher energies there appear a forest of quasi-continuum excitations which set in progressively above the lowest band. 
This dense spectral structure above a certain energy scale is very common in finite unfrustrated
antiferromagnets\cite{Lhuillier,Rousochatzakis_Kagome} and, as we show below, affects the relaxational behavior in a very characteristic way.

We are interested in the damping of the equilibrium fluctuations of the total magnetization $S_z$ 
at not too low $T$ and for $\hbar\omega_e \equiv g\mu_B B \ll J$.  
This damping is triggered by the (phonon-driven) fluctuating portion of various anisotropies.
Here we consider the quadrupolar spin-phonon channel\cite{Villain,Leuenberger,Garanin,NMR_Santini}
which we write in the general form
\be\label{Vsph.Eq}
\mc{V}_{\text{s-ph}}=\sum_{i=1}^{N} \vec{Q}(\vec{s}_i) \cdot \boldsymbol{\Phi}(\vec{r}_i) \equiv  
\sum_{i=1}^{N} \sum_{m=-2}^{2} Q_{i m}^\dagger~\Phi_{i m}~,
\ee
where $\boldsymbol{\Phi}(\vec{r}_i)$ are functions of the local strains or rotation fields, and $Q_m(\vec{s}_i)$ are the quadrupolar operators 
$Q_{\pm 2} (\vec{s})= s_{\pm}^2$, $Q_{\pm 1}(\vec{s}) =  (s_{\pm}s_{z}+s_{z}s_{\pm})/2$, and $Q_0(\vec{s})=  s_z^2$.

It has been found, both experimentally\cite{GSV,NMR_Baek} and numerically\cite{NMR_Santini}, 
that the damping of $S_z$ is mono-exponential (or Markovian) in a large number of nanomagnets. 
The physical origin of this central feature has been shown\cite{Rousochatzakis_PRB} to arise from a dynamical decoupling of $S_z$ 
from the remaining slow degrees of freedom which, in turn, follows from the discreteness of the energy spectrum 
and the conservation law $[S_z,\mc{H}_0]=0$. 
According to the general expression (27) of Ref. [\onlinecite{Rousochatzakis_PRB}], 
$\omega_c$ is given by
\be\label{wc1.Eq}
\omega_c = \frac{\beta}{\chi_0}\sum_{m i, m' i'}\int_0^{\infty} d\omega'~J_{F_{i m} F_{i' m'}^\dagger}(-\omega') 
J_{\Phi_{i m}^\dagger \Phi_{ i' m'}}(\omega')~,   
\ee
where $i\hbar F_{i m} \equiv [S_z,Q_{i m}]=m Q_{i m}$\cite{note1}, 
$\beta=1/k_BT$, and $\chi_0\equiv \beta \langle \delta S_z^2 \rangle$ is the isothermal susceptibility. 
Thus $\omega_c$ is proportional to the frequency overlap between the absorption (-$\omega'<0$) coefficient $J_{F_{i m} F_{i' m'}^\dagger}(-\omega')$ 
and the emission coefficient $J_{\Phi_{i m}^\dagger \Phi_{ i' m'}}(\omega')$ of the host lattice\cite{note2}.
As we show below, this formulation is very fruitful since it allows to identify the frequency regime (or the excitations) which gives the 
dominant contribution to $\omega_c$ in the severe slowing down $T$ regime.

\begin{figure}[!b]
\centering
\includegraphics*[width=0.9\linewidth]{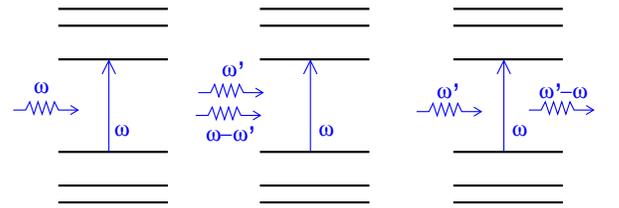}
\caption{(Color Online) Schematic representation of the three lowest order processes which  contribute to $\omega_c$ for any given absorption 
frequency $\omega$ of the nanomagnet. 
(a) The one-phonon absorption process is proportional to the average phonon energy density 
$\omega \rho(\omega)n(\omega)$, where $\rho(\omega)$ is the phonon density of states and 
$n(\omega)=(\exp(\beta\hbar\omega)-1)^{-1}$ is the Bose-Eistein distribution function.
(b) The two-phonon absorption process is proportional to the integral  
$\int d\omega'~\omega'\rho(\omega')n(\omega')(\omega-\omega')\rho(\omega-\omega')n(\omega-\omega')$. 
(c) The inelastic scattering Raman process is proportional to the integral
$\int d\omega'~\omega'\rho(\omega')n(\omega') (\omega'-\omega)\rho(\omega'-\omega) [n(\omega'-\omega)+1]$.}  \label{Processes.Fig}
\end{figure}

To proceed we shall make the reasonable assumption 
that the strain fields (e.g., local librations of the ligand groups) are uncorrelated between different magnetic sites, i.e., 
\be
J_{\Phi_{i m}^\dagger \Phi_{ i' m'}}(\omega')=J_{\Phi_{i m}^\dagger \Phi_{ i m'}}(\omega') \delta_{i i'}
\ee  
(below we drop the site indices since all operators shall refer to a single site).
We further note that the SU(2) invariance of $\mc{H}_0$ at $B=0$ necessitates that
$J_{Q_{m} Q_{m'}^\dagger}(\omega)=J_{Q_{1} Q_{1}^\dagger}(\omega)\delta_{mm'}$. 
Since this remains true at finite $B$ for $k_B T \gtrsim \hbar\omega_e$, we may replace Eq.~(\ref{wc1.Eq}) by
\be\label{wc2.Eq}
\omega_c = \frac{10 N}{\hbar^2} \int_0^{\infty} d\omega'~\frac{J_{s}(-\omega')}{\langle \delta S_z^2 \rangle} 
J_{\overline{\Phi}^\dagger \overline{\Phi} }(\omega')~,   
\ee
where $J_{s}(\omega) \equiv J_{Q_{1} Q_{1}^\dagger}(\omega)$ and 
\be
J_{\overline{\Phi}^\dagger\overline{\Phi}}(\omega)\equiv \sum_{m} m^2 J_{\Phi_m^\dagger \Phi_m}(\omega)/\sum_m m^2~,
\ee 
which defines implicitly an average coupling field $\overline{\Phi}(\vec{r},t)$.
 
For our purposes we use a simple Debye model consisting of 3 acoustic branches with a common sound velocity $c$ 
and a Debye cutoff $\omega_D$ (cf. below), and keep from the strain tensor\cite{Villain,Leuenberger,Garanin} $\boldsymbol{\epsilon}$
only its isotropic (scalar) portion
\be
\eta (\vec{r},t) \equiv \nabla\cdot \vec{u} = \sqrt{ \frac{\hbar}{2 M c} } \sum_{\vec{k}\sigma} \sqrt{k}  
\left( e^{i (\vec{k}\cdot \vec{r}+\omega_{\vec{k}\sigma}t) } a^\dagger_{\vec{k}\sigma} + h.c. \right)
\ee
where $\vec{u}(\vec{r},t)$ denotes the displacement field, 
$M$ is the total mass of the crystal, $\sigma$ runs over the three polarization states and  
$\omega_{\vec{k}\sigma} = c |\vec{k}|$. We may then expand $\overline{\Phi}(\vec{r},t)$ as
\be\label{Phi.Eq}
\overline{\Phi}(\vec{r},t) \simeq v_1 \eta(\vec{r},t) + v_2 \eta^2(\vec{r},t)~,
\ee
where $v_1$ and $v_2$ define two spin-phonon coupling energy parameters whose importance will become clear below.
Since $J_s(\omega)$ is sharply peaked at the Bohr frequencies $\omega_B$ of the cluster, i.e., 
$J_s(\omega)=\sum_{\omega_B} J_s'(\omega_B) \delta(\omega-\omega_B)$, we may rewrite Eq.~(\ref{wc2.Eq}) as
\be\label{wc3.Eq}
\frac{\omega_c}{10 N} = \sum_{\omega_B>0} \frac{J_s'(-\omega_B)}{\langle \delta S_z^2 \rangle}
\left( \frac{v_1^2}{\hbar^2} J_{\eta\eta}(\omega_B) +\frac{v_2^2}{\hbar^2} J_{\eta^{2}\eta^2}(\omega_B) \right)~.   
\ee
Using the statistical factor  $n(\omega)=(e^{\beta\hbar\omega}-1)^{-1}$ and the mass density $\rho_m$, we have 
for $0<\omega<\omega_D$:  
\be
J_{\eta\eta}(\omega)=\frac{3}{2\pi}\frac{\hbar}{\rho_m c^5}  n(\omega) \omega^3\equiv J_{1ph}(\omega)
\ee 
which is the contribution from direct one-phonon processes, while $J_{\eta^{2}\eta^2}(\omega)$   
is the sum of the two-phonon and Raman contribution,  given respectively by  
\begin{eqnarray}\label{J2ph.Eq}
J_{2ph}(\omega)&=&A\int_0^{\omega} d\omega'\omega'^3 n(\omega') (\omega-\omega')^3 n(\omega-\omega')\nonumber\\ 
J_{R}(\omega)&=&2A\int_{\omega}^{\omega_D+\omega} d\omega'\omega'^3 n(\omega')\times \nonumber\\
&&(\omega'-\omega)^3\left[ n(\omega'-\omega)+1\right]~, 
\end{eqnarray} 
where $A\equiv\frac{9}{4\pi^3} \frac{\hbar^2}{\rho_m^2 c^{10}}$.
These processes are represented schematically in Fig. \ref{Processes.Fig}.

Equations (\ref{wc3.Eq}-\ref{J2ph.Eq}) are the starting point of our calculations for $\omega_c$. 
The quadrupolar spectral density $J_s(\omega)$ is obtained by a full thermodynamic calculation using exact diagonalizations.

\begin{figure}[!t]
\centering
\includegraphics*[width=0.8\linewidth]{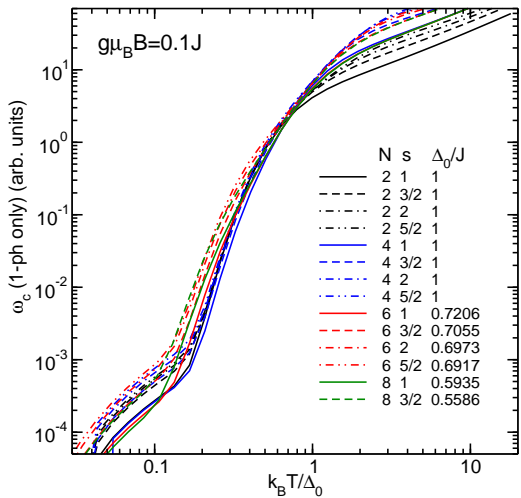}
\caption{(Color Online) Near-universal behavior of $\omega_c$ vs. $k_BT/\Delta_0$ for rings with different $N$ and $s$.
The dramatic slowing-down (by 3-4 decades) takes place in the regime $0.2\Delta_0\lesssim k_BT\lesssim 2\Delta_0$, 
where $\omega_c$ follows Eq.~(\ref{wc4.Eq}). 
The limits of this distinctive $T$ regime can be understood by looking at Fig. \ref{Overlap.Fig}:
It begins when the one-phonon spectral peak starts overlapping the quasi-continuum portion of the quadrupolar density, 
and terminates when it has sampled its entire bandwidth.  
At $k_BT\gtrsim 2 \Delta_0$, $\omega_c \propto T$ (Raman processes or high-frequency phonons alter this behavior), 
while at very low $T$, $\omega_c \sim \omega_e^3 /(e^{\beta\hbar\omega_e}-1)$ (cf. text).}  
\label{All_N_s.Fig}
\end{figure}

\section{Slowing down effect and near-universal power-law scaling}
\begin{figure}[!t]
\centering
\includegraphics*[width=1\linewidth]{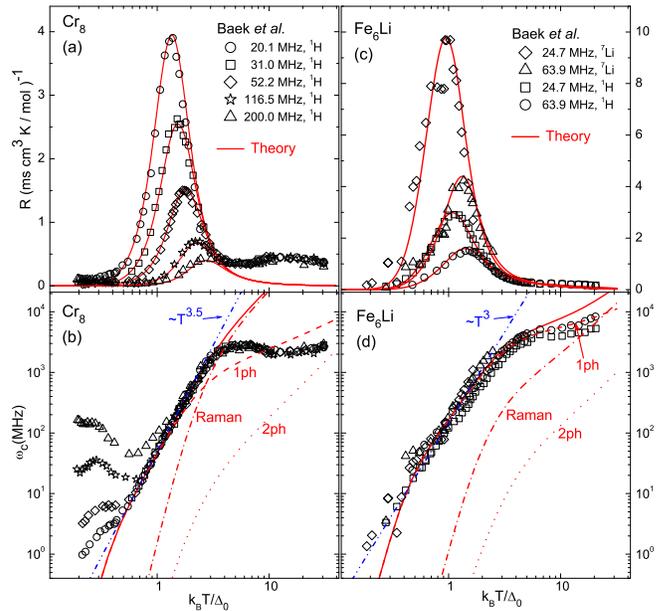}
\caption{(Color Online) Comparison between theory (solid red lines) and experimental data (symbols) for $R$ [cf. Eq.~(\ref{R.Eq})] and $\omega_c$ 
for Cr$_8$ (left) and Fe$_6$Li (right). The one-phonon (dashed), two-phonon (dotted) and Raman (dash-dot) contributions to $\omega_c$ are also shown, along with 
the respective power-law scalings (dash-double dot, blue lines) for $0.2 \Delta_0\lesssim k_BT\lesssim 2 \Delta_0$.} 
\label{Cr8.Fig}
\end{figure}

Figure \ref{Overlap.Fig} shows $J_s(-\omega)$,  $J_{1ph}(\omega)$, $J_{2ph}(\omega)$ and $J_{R}(\omega)$ for Cr$_8$ 
at $\hbar\omega_e = 0.066 \Delta_0$ and for $k_BT = 0.19 \Delta_0$ (a) and $2.22 \Delta_0$ (b). The values of the parameters used here 
are the ones that fit the $1/T_1$ data (cf. Table~\ref{Table}).
Apart from identifying a number of excitation lines such as L$_1$ (transitions within the lowest triplet), 
L$_2$ (transitions from the lowest triplet to the lowest quintet) and L$_3$ (transitions from the lowest singlet to the lowest quintet)\cite{note3},
we find a number of general features which originate in the overall spectral structure of AFMR's described above. 
We first emphasize the increasingly gapped structure at low $\omega$ and $T$, and the fact that $J_s(-\omega)$ has essentially no weight below the line L$_2$. 
This structure is altered very quickly by thermal excitations since the latter increase the number of available resonant channels 
and thus give rise to a dense fluctuation spectrum at higher $T$.  Importantly,  $J_s(-\omega)$ remains appreciable over an overall bandwidth 
$\hbar\omega_{\text{max}}\sim 10-20 \Delta_0$.  
Given now the behavior of the phononic density (cf. Fig.~\ref{Overlap.Fig})  the relaxation process can be understood as follows. 
At $k_BT\lesssim \hbar\omega_e \ll \Delta_0$ the system is essentially opaque to the available (thermally excited) phonons since 
there is no appreciable overlap between $J_s(-\omega)$ and $J_{1ph}(\omega)$. 
Only the Zeeman line L$_1$ contributes, giving $\omega_c \sim \omega_e^3 /(e^{\beta\hbar\omega_e}-1)$.
However, as soon as  the one-phonon spectral peak reaches the quasi-continuum regime (above the line L$_2$) 
a significant number of spin-phonon resonant channels are quickly thermally activated. In fact, Fig.~\ref{Overlap.Fig}(b) shows that 
$J_{1ph}(\omega)$ samples almost the entire electronic spectral weight up to $\omega_{\text{max}}$ already at $k_B T \sim 2 \Delta_0$. 
This marks the existence of a special regime $0.2 \Delta_0 \lesssim k_BT\lesssim 2 \Delta_0$, where a dramatic change 
of $\omega_c$  by 3-4 decades (cf. Fig.~\ref{All_N_s.Fig}) takes place. 
We emphasize here that the relaxation process in this $T$ regime is driven by the quasi-continuum portion 
of the quadrupolar spectrum and is minimally affected by the lowest excitation lines. 
This shows that $\omega_c$ is not a nearly exponential function in $1/T$ in this regime.
In fact, we may go one step further and inquire into the observed power-law scaling by employing a 
steepest-descent expansion in Eq.~(\ref{wc2.Eq}) which relies on two central ingredients: 
(i) the dominant contribution to the overlap comes from the regime around the one-phonon peak 
$\hbar\omega_p(T)\simeq 2.82 k_BT$ (note that Fig.~\ref{Overlap.Fig} is in logarithmic scale), 
and (ii) the quasi-continuum character of $J_s(-\omega)$ in the respective $T$ and $\omega$ regime. 
To leading order 
\be\label{wc4.Eq}
\omega_c \propto T^4 f(T) 
\ee 
where $f(T)\equiv J_s(-\omega_p(T),T)/\chi_0T$, and the strong $T^4$ pre-factor emerges from the functional form of $J_{1ph}(\omega,T)$ alone.  
The function $f(T)$ varies with $N$ and $s$ but shows always a much weaker (sub-linear) $T$ dependence. 
This can be seen by our calculations shown in Fig.~\ref{Cr8.Fig} (cf. text below) which are in excellent agreement with experimental data. 
Hence, the $T^{3.5\pm 0.5}$ scaling law reported in Ref. [\onlinecite{NMR_Baek}] is largely due to the above $T^4$ leading pre-factor. 
It is thus essentially a characteristic fingerprint of one-phonon acoustic processes and of
the quasi-continuum nature of the electronic quadrupolar spectrum at high energies.

All of the above features are demonstrated in Fig.~\ref{All_N_s.Fig} which shows the one-phonon contribution to $\omega_c$ 
for various $N$ and $s$. Indeed, one identifies a distinctive regime $0.2 \Delta_0\lesssim k_BT\lesssim 2 \Delta_0$, 
where $\omega_c$ shows a dramatic drop by 3-4 decades.
Figure \ref{All_N_s.Fig} explains yet another of the central findings of Baek \textit{et al.}\cite{NMR_Baek}, namely that all curves
fall almost on top of each other when plotted against $k_BT/\Delta_0$. 
This is clearly due to the overall similar spectral structure of AFMR's and, in particular, 
due to the fact that the forest of quasi-continuum excitations sets in at an energy scale fixed by $\Delta_0\simeq 4 J/N$ (and not by $J$). 

We should add here that the details of the low-energy spectrum such as the character of the lowest L band or the spin-wave E band\cite{Waldmann} 
do not play any special role compared to other excitations. 
The quasi-continuum portion of the spectrum contains excitations from and towards both the L and E bands,
but it is the global dense aspect of the spectrum that matters.

\section{Comparison to Nuclear spin-lattice relaxation rate data}
Let us now describe our calculations for $1/T_1$ and compare to experimental data for Fe$_6$Li, Fe$_6$Na and Cr$_8$. 
We begin with the expression\cite{NMR_Baek,NMR_Santini,mythesis} 
\be\label{T1.eq}
1/T_1=\mc{A}_{zz} J_{s_{0z}s_{0z}}(\omega_L)=\frac{\mc{A}_{zz}}{N^2} J_{S_{z}S_{z}}(\omega_L)
\ee 
where $\vec{s}_0$ is the ionic spin with the shortest distance $r_0$ from the nuclear spin,  
and $\mc{A}_{zz}=\hbar^2 \gamma_e^2\gamma_n^2/r_0^6$ is the corresponding hyperfine amplitude\cite{note4}. 
Using\cite{Rousochatzakis_PRB} $J_{S_zS_z}(\omega_L)=2\langle\delta S_z^2 \rangle \frac{\omega_c}{\omega_c^2+\omega_L^2}$,
and $\chi_{\text{mol}}=N_A g^2 \mu_B^2 \beta \langle\delta S_z^2 \rangle$ (where $N_A$ is Avogadro's number) we get 
\be\label{R.Eq}
R \equiv \frac{1}{T_1\chi_{\text{mol}}T} = 4585.3 ~ \frac{\gamma_n^2}{N^2 r_0^6}~\frac{\omega_c}{\omega_c^2+\omega_L^2}~,
\ee 
where $1/T_1$ is given in ms, $\gamma_n$ in MHz/T, $\chi_{\text{mol}}T$ in cm$^3$ K/mol, $r_0$ in $\AA$, and $\omega_{c,L}$ in MHz. 
According to Eq.~(\ref{R.Eq}),  the fit of $r_0$ is controlled by the magnitude of $1/T_1$,  
while the values of $v_1$ and $v_2$ of Eq.~(\ref{wc3.Eq}) can be found by adjusting the position and the width of the $1/T_1$ peak. 
We should note however that $v_2$ affects $\omega_c$ only at the high-$T$ side of the peak where high energy phonons start to play a role 
and thus its estimate is generally less accurate. 

Figure \ref{Cr8.Fig} shows our optimal fits to the data of  $R$ and $\omega_c$ for Cr$_8$ and Fe$_6$Li,
along with the separate contributions of direct, two-phonon and Raman processes. Similar fits (not shown here) are obtained for  Fe$_6$Na. 
The corresponding estimates for $r_0$, $v_1$ and $v_2$ are given in Table~\ref{Table} and are of the correct order of magnitude. 
As to the value of the Debye cutoff, we find that our fits remain very good for $\hbar\omega_D \sim 10 J$\cite{note5},
which is consistent with the reported estimates of the Debye temperature $\Theta_D$ [cf. Table~\ref{Table}].
The agreement at intermediate $T$ where the enhancement of $1/T_1$ takes place is remarkably good.
In particular, we find that the relaxation is dominated by one-phonon processes up to $k_BT\sim 2\Delta_0$ 
for Cr$_8$ ($\Delta_0/k_B\simeq 9.607$K) and up to $\sim 6 \Delta_0$ for Fe$_6$Li ($\Delta_0/k_B\simeq 14.526$K). 
On the other hand, the power-law scaling (dashed-double dot blue lines) is valid up to $\sim 2 \Delta_0$ for both clusters.
We also find that our model does not account for the behavior at very high $T$. This can be either due to another contribution to $1/T_1$
(e.g., electronic $T_2$ processes\cite{Spanu,Pilawa,mythesis} not included in Eq.~(\ref{R.Eq})) or due to the actual details 
of the high-energy phonon modes which should play a role at high $T$.
Finally, the discrepancy at very low-$T$ in Cr$_8$ arises from an additional peak in $1/T_1$ which is currently not understood.
\begin{table}[!t]
\caption{Known data\cite{SynthesisData,SpecificHeatData} (first three columns) and fitting parameters for Fe$_6$Li, Fe$_6$Na, and Cr$_8$.
Here $\tilde{c}\equiv \frac{c[\text{cm/s}]}{2\cdot 10^{5}}$.}
\label{Table} 
\begin{ruledtabular}
\begin{tabular}{l|c|c|c|c|c|l}
Cluster & $\rho_m$(g/cm$^3$) & $\Theta_D$(K) & $J$(K) & $v_1/\tilde{c}^{5/2}$(K) & $v_2/\tilde{c}^5$(K) & $r_0$(\AA) \\
\hline
Fe$_6$Li &1.45  & 217.8         & 21    &  0.498 & 2.861    & 4.59 \\
Fe$_6$Na &1.42 & 209.8          & 28    &  0.200 & 1.277  & 4.16  \\
Cr$_8$  &1.08  & 154$\pm $10 & 17.2 & 1.124 & 14.376    & 4.10\\
\end{tabular}
\end{ruledtabular}
\end{table}

\section{Conclusions}
We have presented a microscopic theory which identifies the mechanism responsible for the dramatic 
slowing down effect observed in antiferromagnetic wheels. 
Our central key result is that in this class of nanomagnets  the relaxation 
is driven by the quasi-continuum portion of the quadrupolar spectrum and not by the low-lying excitations. 
This is at the heart of the intriguing power-law $T$ dependence of the electronic correlation frequency $\omega_c$
with an exponent close to 4. The onset of this scaling is fixed by the lowest spin gap $\Delta_0\simeq 4J/N$ (and not by $J$) and 
this explains why experimental curves of $\omega_c$ for different cluster sizes and spins nearly coincide when $T$ is rescaled by $\Delta_0$. 
Since the spectral structure of AFMR's is very common in finite-size antiferromagnets most of the above 
qualitative features must carry over to other antiferromagnetic clusters as well.
Hence we believe that the present slowing down mechanism of AFMR's is common in the more general class of 
antiferromagnetic clusters. 
More generally, the present theory can be applied to the majority of clusters reported in the literature for $T\gtrsim 1$ K. 
It can also be extended to clusters with $s=1/2$ by using an appropriate relaxation channel.\cite{note0} 
It is thus our hope that it will be valuable for further investigations on the understanding and characterization of 
relaxation mechanisms in magnetic nanoclusters. 

\acknowledgments{We thank F. Mila and M. Belesi for fruitful discussions. 
The work at EPFL was supported by the Swiss National Fund.
Work at the Ames Laboratory was supported by the Department of  
Energy - Basic Energy Sciences under Contract No. DE-AC02-07CH11358.}


\begin{thebibliography}{99}
\bibitem{GSV} D. Gatteschi, R. Sessoli, and J. Villain, \textit{Molecular Nanomagnets} (Oxford University Press, Oxford, 2006), and references therein.
\bibitem{NMR_Santini} P. Santini, S. Carretta, E. Liviotti, G. Amoretti, P. Carretta, M. Filibian, A. Lascialfari, and E. Micotti, 
\prl \textbf{94}, 077203 (2005).
\bibitem{Rousochatzakis_PRB} I. Rousochatzakis, \prb {\bf 76}, 214431 (2007).
\bibitem{Ni10}  S. Carretta, P. Santini, G. Amoretti, M. Affronte, A. Candini, A. Ghirri, I. S. Tidmarsh, R. H. Laye, R. Shaw, and E. J. L. McInnes, \prl {\bf 97}, 207201 (2006). 
\bibitem{NMR_Baek} S. H. Baek, M. Luban, A. Lascialfari, E. Micotti, Y. Furukawa, F. Borsa, J. van Slageren, and A. Cornia, 
\prb \textbf{70}, 134434 (2004).
\bibitem{note0} Magnetic clusters with $s=1/2$ must be treated separately since they do not have a quadrupole moment and thus
a different relaxation channel (e.g. dipolar channel or fluctuating Dzyaloshinskii-Moriya interactions) must be invoked. 
\bibitem{SchnackLuban} J. Schnack and M. Luban, \prb{\bf 63}, 014418 (2000).
\bibitem{Waldmann} O. Waldmann, \prb{\bf 65}, 024424 (2002).
\bibitem{Larry}  L. Engelhardt and M. Luban, \prb {\bf 73}, 054430 (2006). 
\bibitem{Lhuillier} C. Lhuillier, cond-mat/0502464.
\bibitem{Rousochatzakis_Kagome} For frustrated clusters see e.g., I. Rousochatzakis, A. M. L\"auchli, and F. Mila, \prb{\bf 77}, 094420 (2008).
\bibitem{Villain} J. Villain, F. Hartman-Boutron, R. Sessoli, and A. Rettori, Europhys. Lett. {\bf 27}, 159 (1994); 
F. Hartmann-Boutron, P. Politi, and J. Villain, Int. J. Mod. Phys. B {\bf 10}, 2577 (1996). 
\bibitem{Leuenberger} M. N. Leuenberger, and D. Loss, Phys. Rev. B {\bf 61}, 1286 (2000). 
\bibitem{Garanin} E. M. Chudnovsky, D. A. Garanin, and R. Schilling, Phys. Rev. B \textbf{72}, 094426 (2005).
\bibitem{note1} As expected, only the $m\neq 0$ components of $\vec{Q}$ can drive the relaxation of $S_z$. 
\bibitem{note2} Here, the spectral density for any pair of operators $A$ and $B$ is defined as 
$J_{A B}(\omega)=\int_{-\infty}^{\infty} dt e^{i\omega t} \langle A(0) B(t) \rangle$.    
\bibitem{note3} Note that there is no absorption line at $\Delta_0\pm \hbar\omega_e$ since transitions from the ground state 
to the lowest triplet are forbidden by symmetry for our quadrupolar spin-phonon coupling. 
\bibitem{mythesis} I.~Rousochatzakis, Ph. D. thesis, Iowa State University (2005).
\bibitem{note4} For $^7$Li NMR in Fe$_6$Li, one must multiply Eq. (\ref{T1.eq}) with an extra factor of $N$ 
since the nuclear spin resides at the center of the ring and thus all ions contribute equally to the nuclear relaxation.
\bibitem{SynthesisData} G. L. Abbati \textit{et al.}, Inorg. Chem. {\bf 1997}, 36, 6443-6446;  A. Caneschi  \textit{et al}, 
Angew. Chem. Int. Ed. Engl.  {\bf 1995}, 34, 467; J. van Slageren \textit{et al}, Chem. Eur. J. {\bf 2002}, 8, 277; 
\bibitem{SpecificHeatData} M. Affronte, J. C. Lasjaunias, and A. Cornia, Eur. Phys. J. B {\bf 15}, 633-639 (2000); 
M. Affronte, T. Guidi, R. Caciuffo, S. Carretta, G. Amoretti, J. Hinderer, I. Sheikin, A. G. M. Jansen, A. A. Smith, R. E. P. Winpenny, J. van Slageren, and D. Gatteschi, Phys. Rev. B {\bf 68}, 104403 (2003); 
\bibitem{note5} This is an approximate lower bound of $\omega_D$ since, as can be understood from Fig.~\ref{Overlap.Fig}, 
larger values do not alter $\omega_c$ quantitatively at intermediate $T$.
\bibitem{Spanu} L. Spanu and A. Parola, \prb{\bf 72}, 212402 (2005).
\bibitem{Pilawa} B. Pilawa, R. Boffinger, I. Keilhauer, R. Leppin, I. Odenwald, W. Wendl, C. Berthier, and M. Horvati\'c, 
\prb{\bf 71}, 184419 (2005).
\end{thebibliography}
\end{document}